\documentclass[conference]{IEEEtran}
\IEEEoverridecommandlockouts

\usepackage{cite}
\usepackage{amsmath,amssymb,amsfonts}
\usepackage{algorithmic}
\usepackage{graphicx}
\usepackage{subfigure}
\usepackage{textcomp}
\usepackage{xcolor}
\usepackage{multirow}
\usepackage{enumitem}
\def\BibTeX{{\rm B\kern-.05em{\sc i\kern-.025em b}\kern-.08em
    T\kern-.1667em\lower.7ex\hbox{E}\kern-.125emX}}
\begin{document}
\title{Using Machine Learning to Automate Mammogram Images Analysis}
\author{
  Xuejiao Tang$^1$, Liuhua Zhang$^2$, Wenbin Zhang$^3$, Xin Huang$^3$\\ Vasileios Iosifidis$^1$, Zhen Liu$^4$, Mingli Zhang$^5$, Enza Messina$^6$ and Ji Zhang$^7$\\
  
  $^1$Leibniz University Hannover, Germany $^2$Memorial University of Newfoundland, Canada \\
  $^3$University of Maryland, Baltimore County, USA $^4$Guangdong Pharmaceutical University, China\\
  $^5$McGill University, Canada 
  $^6$University of Milano-Bicocca, Italy
  $^7$University of Southern Queensland, Australia\\

  xuejiao.tang@stud.uni-hannover.de, lz4038@mun.ca, \{wenbinzhang,xinh1\}@umbc.edu, iosifidis@l3s.de\\
   liu.zhen@gdpu.edu.cn, mingli.zhang@mcgill.ca, enza.messina@unimib.it, ji.zhang@usq.edu.au
  }
\maketitle
\begin{abstract}
Breast cancer is the second leading cause of cancer-related death after lung cancer in women. Early detection of breast cancer in X-ray mammography is believed to have effectively reduced the mortality rate since 1989. However, a relatively high false positive rate and a low specificity in mammography technology still exist. In this work, a computer-aided automatic mammogram analysis system is proposed to process the mammogram images and automatically discriminate them as either normal or cancerous, consisting of three consecutive image processing, feature selection, and image classification stages. In designing the system, the discrete wavelet transforms (Daubechies 2, Daubechies 4, and Biorthogonal 6.8) and the Fourier cosine transform were first used to parse the mammogram images and extract statistical features. Then, an entropy-based feature selection method was implemented to reduce the number of features. Finally, different pattern recognition methods (including the Back-propagation Network, the Linear Discriminant Analysis, and the Naive Bayes Classifier) and a voting classification scheme were employed. The performance of each classification strategy was evaluated for sensitivity, specificity, and accuracy and for general performance using the Receiver Operating Curve. Our method is validated on the dataset from the Eastern Health in Newfoundland and Labrador of Canada. The experimental results demonstrated that the proposed automatic mammogram analysis system could effectively improve the classification performances.
\end{abstract}


\section{Introduction}
Breast cancer is the most commonly diagnosed form of cancer in women and the second-leading cause of cancer-related death after lung cancer~\cite{b1}. Statistics from the 
American Cancer Society indicate that approximately 232,670 (29\% of all cancer cases) American women will be diagnosed with breast cancer, and an estimated 40,000 (15\% of all cancer cases) women will die of it in 2014 ~\cite{b2}. In other words, 637 American women will be diagnosed with breast cancer, and 109 women will die of it 
every day. Similar statistics were also found in Canada, where approximately 23,800 (26\%) women were diagnosed with breast cancer, and 5,000 (14\%) died from it in 2013~\cite{b3}. Under this circumstance, detection and diagnosis of breast cancer has already drawn a great deal of attention from the medical world.

Studies show that early detection, diagnosis and therapy is particularly important to prolong lives and treat cancers~\cite{b4}. If breast cancer is found early, the five-year survival rate of patients in stage 1 could reach 90\% with effective treatment. Medical imaging technology is one of the main methods for breast cancer detection. Commonly used medical imaging technologies include X-ray mammography, Computer Tomography (CT), ultrasound and Magnetic Resonance Imaging (MRI), Positron Emission Tomography (PET), and Single-Photon Emission Computed Tomography (SPECT). Among these technologies, mammography achieves the best results in early detection of asymptomatic breast cancer and is one of the least expensive ones. For this reason, it has become the principal method of breast cancer detection in clinical practice, and one of the most effective ways for general breast cancer survey. 

Masses and calcifications (including marcocalcifications and microcalcifications) are the most common and basic symptoms of breast cancer. Masses in mammography can be recognized as a local, high-contrast area, but the value of contrast is not unique. It changes when imaging conditions, sizes and backgrounds change. The X-ray absorption rates of masses are very close to dense glandular tissue in breast and other dense tissues. In addition, the boundaries of masses are always mixed with background structures. 
Therefore, microcalcification detection remains one of the most popular topics in medical image processing research~\cite{b5}.

Modern equipment has improved the technical aspects of mammography. However, there still exist a relatively high false positive rate and a low specificity, due to fundamental physical limitations such as unobvious lesions, as well as controllable factors like radiologists’ inexperience in reading mammograms. This latter issue has been addressed using double reading, where two radiologists make their own judgments independently based on the same mammogram, and then combine and discuss both opinions. However, the solution is expensive and highly relying on radiologists' experience. To improve the accuracy of reading mammograms, interest in Computer-Aided Diagnosis (CADx) solutions has emerged~\cite{b4}.

In this paper, we focus on breast cancer detection using a computer-aided automatic mammogram analysis system to improve the accuracy of diagnosis. In proposed analysis system, entropy-based method was implemented for feature selection as well as different pattern recognition methods, including the Back-propagation (BP) Network, the Linear Discriminant Analysis (LDA), the Naive Bayes (NB) Classifier, and voting schemes were employed for image classification.

The rest of the paper is organized as follows. Section~\ref{related work} presents background regarding breast cancer detection and relevant work. Section~\ref{system design} proposes the automatic mammogram images analysis system. Subsequently, the experimental results are presented and analyzed in Section~\ref{experiments}. Finally, Section~\ref{conclusions} draws the conclusion and discusses lines of future work.

\section{RELATED WORK}
\label{related work}

Calcification detection and classification has been an important research target of the automatic mammogram analysis systems. A wide variety of approaches have so far been developed to improve the detection performance. However, it's still challenging for the following reasons: 1) microcalcifications occur in various sizes, shapes, and distributions; 2) microcalcifications have low contrast in the region of interest (ROI); 3) dense tissue and/or skin thickness make suspicious lesion areas difficult to detect (especially in young women); 4) the dense tissue is easily misunderstood as microcalcification, which results in high false positive rates among most existing algorithms. 

Feature selection is crucial for detection and classification. A number of efforts have been devoted to employing numerous different features in the application of mammogram analysis. For example, the geometrical and statistical features in~\cite{b17,b18} are used for classification. But these methods depend on the image translation, scaling and rotation heavily, which results in misclassification. Consequently, these approaches can be improved by using a robust factor, such as Fourier transform, which is certainly translation invariant~\cite{b19}. In the data time-frequency analysis, the Fourier transform is traditionally applied, which is a global transform between the time and frequency domain. Therefore, the Fourier transform cannot express the local properties of signals in the time and frequency domains simultaneously. However, these local properties are the key characteristics of non-stationary signals in some circumstances. To analyze and process non-stationary signals, wavelet transforms is one of the methods to retain important temporal information in the frequency domain or partial frequency information in the time domain. In this work, both the Fourier transform and wavelet transforms are applied at the data transform stage to extract the features. However, not all of the extracted features benefit classification performance. For a feature to be useful in classification, it should be closely and uniquely associated with a certain class~\cite{b13}. Ideally, the feature will correlate with the desired class independent of the presence of other classes. If these conditions are met, the feature reduction (selection) problem can be addressed by measuring the correlation with that class then establishing a pass threshold. The pass threshold eliminates features that correlate poorly. There are two common approaches used to measure the correlation between two random variables, in this case between feature and class~\cite{b14}. The first is the linear correlation, where the variation in a feature value is compared to the variation in class value. The second approach and the one adopted for our study is Information Gain, a concept based on the reduction of entropy in the dataset.

A target range for the number of features was determined from the work of, Lei and Huan~\cite{b15}, they proposed a fast correlation based filter approach and conducted an efficient way of analyzing feature redundancy. Their new feature selection algorithm was implemented and evaluated through extensive experiments comparing with other related feature selection algorithms based on ten different kinds of feature types. The number of features ranged from 57 to 650, and the sample size of feature types ranged from 32 to 9338. At the end of the experiment, they recorded the running time of the proposed system and the number of features selected for each algorithm. The results showed that the average selected number of features was 15 for the five compared feature selection algorithms, and the selected features could lead to classification accuracy to around 89\%. In this research, we chose a threshold of information gain which could lead to around 15 features left. Entropy is a measure of the uncertainty of a random variable~\cite{b15}. 
\section{System Design}
\label{system design}
\begin{figure}[!htbp]
\vspace{-0.2cm}
\centerline{\includegraphics[width = 8cm]{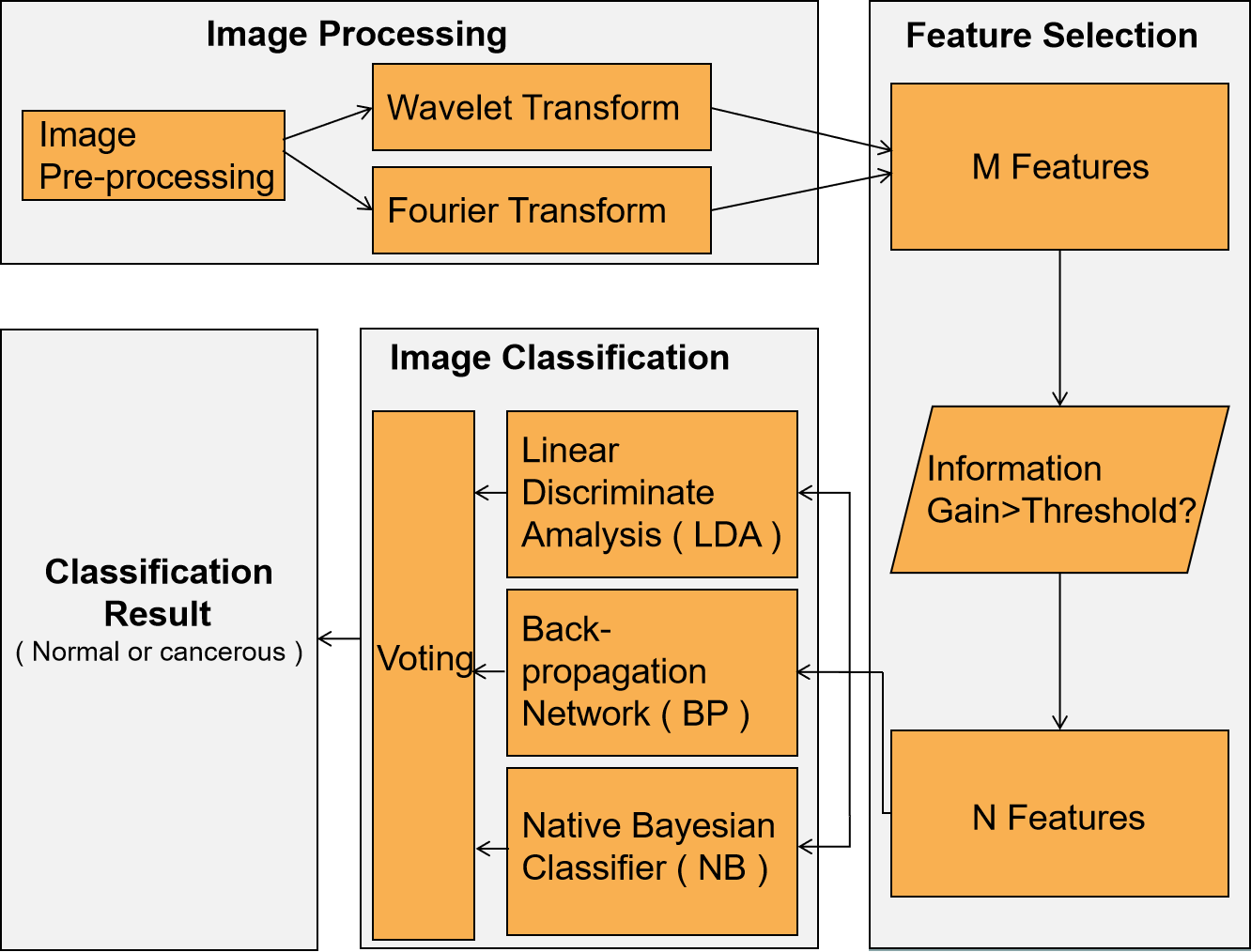}}
\vspace{-0.2cm}
\caption{Framework of automatic mammogram analysis system.}
\label{fig:framework}
\vspace{-0.4cm}
\end{figure}

The proposed automatic mammogram analysis system comprised of three consecutive stages, including the image processing stage, the feature selection stage and the image classification stage. Fig.~\ref{fig:framework} visualizes the framework of the proposed system with each components being detailed in the following sections.

\subsection{Image Processing}
\label{sys:img process}
In the first image processing stage, a set of scalar features were extracted from an original image. This stage consisted of two steps: image pre-processing and data transform (including wavelet and Fourier transforms). In the image pre-processing step, the original digitized mammogram image is flipped, de-noised, and scaled to a common maximum value. In the data transform step, the normalized images are decomposed by three wavelet transforms with different bases (Daubechies db2, Daubechies db4, and Biorthogonal bior6.8) and the Fourier transform separately.

Multiple levels of decomposition were used, and four images are produced at each 
level of the decomposition. Finally, four statistic features, including the mean, 
standard deviation, skewness and kurtosis of the image intensities, were calculated.

\subsubsection{Mammogram Image Pre-processing}

for the automatic mammogram analysis system, the original images are different 
in size and directions. Furthermore, artefacts and noise may also exist in some 
mammograms, which would generate wrong or poor analysis result. Thus, several 
mammogram pre-processing steps were implemented to regularize the appearance of 
the images, and remove unnecessary artefacts and noise. Based on the studies~\cite{b20}, the steps taken in this work are as below:

\begin{figure}[!htbp]
\vspace{-0.3cm}
\centering
\subfigure{
\includegraphics[height = 3.2cm]{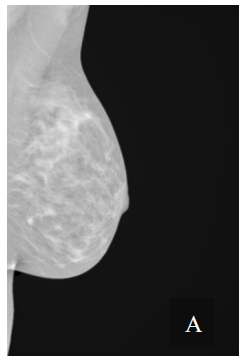}
}
\subfigure{
\includegraphics[height = 3.2cm]{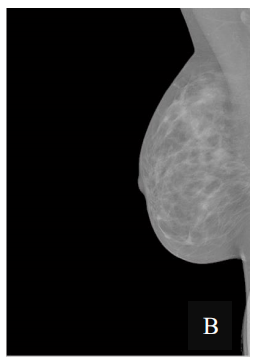}
}
\subfigure{
\includegraphics[height = 3.2cm]{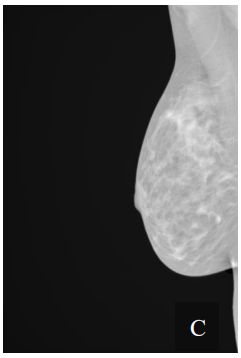}
}
\vspace{-0.2cm}
\caption[An example of MLO view mammogram: A. Right side; B. Left 
side; C. the image after orientation matching of A.]
{An example of MLO view mammogram: A.Right side; B.Left side; C.the image after orientation matching of A.}
\label{fig:example_MLO}
\vspace{-0.4cm}
\end{figure}

\begin{enumerate}[labelsep = .5em, leftmargin = 0pt, itemindent = 3em]
\item[i)] {Orientation Matching:} our study only involved the MLO mammogram presentation. In these, the right and left breasts point to the opposite sides in the mammogram image. Therefore, flipping one of the breasts to the same direction as the other one ensures that all images pointed in the same direction, preventing changes in the wavelet transform coefficients due only to the directionality change between right and left images. The sharp edge between the tissue and the dark background is a major feature in all images that affects this change. As shown in Fig.~\ref{fig:example_MLO}, the intensity of right breast images falls from left to right across this edge, while it rises in left breast images. This would change the sign of the calculated wavelet coefficient. Fig.~\ref{fig:example_MLO} shows the result of orientation matching of an example of Medial Lateral Oblique (MLO) view mammogram. Fig.~\ref{fig:example_MLO} A and B respectively show the right and left breast images of a patient with tiny microcalcifications in her breast tissue. Fig.~\ref{fig:example_MLO} C shows the reflected image of orientation matching of the right breast.

\begin{figure}[!htbp]
\centering
\subfigure{
\includegraphics[height = 3.0cm]{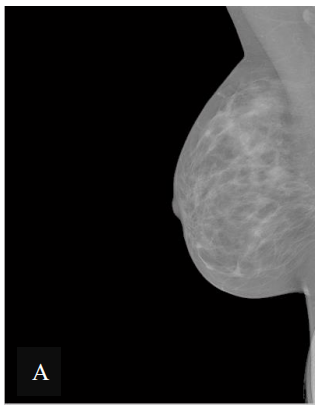}
}
\subfigure{
\includegraphics[height = 3.0cm]{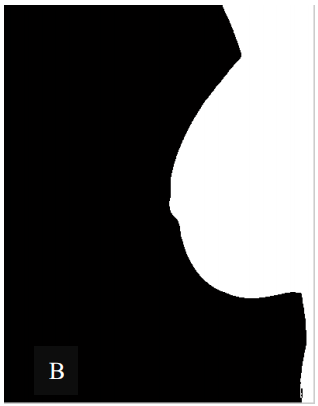}
}
\caption[A. Mammogram image before background thresholding; B. The 
thresholded binary image used to mask the original image.]
{A. Mammogram image before background thresholding; B. The 
thresholded binary image used to mask the original image.}
\label{fig: thresholded binary image}
\end{figure}

\item[ii)] {Background Thresholding:} signal outside the tissue is non-informative and was removed by binary masking. Concretely, a threshold is set to create binary images. Pixels with lower intensity value than the threshold are set to zero. A satisfactory threshold can remove all irrelevant information in the background pixels, and leave foreground objects unaltered. One of the most commonly used method to choose the threshold 
is Otsu’s Method~\cite{b12}, which assumes that the image to be thresholded contains two 
classes of pixels or bi-modal histogram (e.g. foreground and background). The method
then calculates the optimum threshold separating those two classes so that their 
combined spread (intra-class variance) is minimal~\cite{b12}. It also assumes that the 
foreground and background intensities are normally distributed, and it chooses the 
threshold level which minimizes the segmentation error between the two regions. The attenuation of x-rays passing through the tissue affects the intensity in the images, and is influenced by the thickness and density of the tissue. Therefore, tissue pixels which fall below the conservative threshold are predominantly from the edges 
of the tissue region where the breast tissue is thin and uncompressed. While a few pixel layers may be removed by this method, it was deemed acceptable as any pathology that exists this close to the surface of a patient’s skin should be readily detectable by conventional examination without the aid of mammography. In this work, the binary thresholding, which sets all pixels below a threshold, was set to an intensity of zero and all pixels above the threshold to an intensity of one (see Fig.~\ref{fig: thresholded binary image}). The output image of the process is the pixel-by-pixel product of the binary mask image and the original image. In this way, all background pixels of the output image are set to zero of intensity, while all foreground pixels are unaffected.
\begin{figure}[!htbp]
\centering
\subfigure{
\includegraphics[height = 3.0cm]{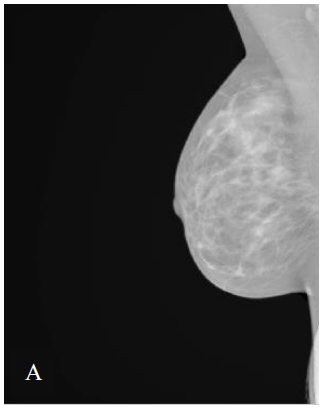}
}
\subfigure{
\includegraphics[height = 3.0cm]{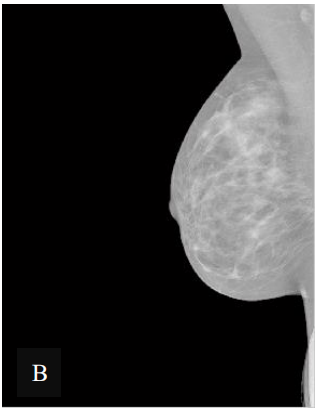}
}
\caption[Mammogram image before A and after B intensity matching.]
{Mammogram image before A and after B intensity matching.}
\label{fig:intensity matching}
\end{figure}
\item[iii)] {Intensity Matching:} intensity matching is the last pre-processing step applied to the images before they are ready for data transforms. In this step, all mammograms are linearly scaled to an
intensity of 0.0 to 1.0. This intensity matching process can be defined by
\begin{equation}
\label{equ:Intensity Matching}
    img\_out=\frac{img\_in}{max(img\_in)}
\end{equation}
where $img\_in$ is the input image following the background thresholding step, and $img\_out$ is the intensity-matched image whose pixel intensities range from zero to one. This step ensures the uniformity across all different mammogram images, because their pixel intensities ranges could differ with machines settings. It can be seen in Fig.~\ref{fig:intensity matching} that there is tiny difference before and after the intensity matching procedure. The broader spread in intensities would increase the variations in different tissue types and densities. (the maximum relative intensity prior to normalization was 0.92).
\end{enumerate}

\subsubsection{Data Transforms}
once the images are pre-processed to minimize the differences between images that were not related to differences in the physical composition of the breast tissue, the wavelet and Fourier transforms were performed on the images. The images were all sampled to 1024×1024 pixels, which would allow maximum 10 levels of decomposition, since dyadic sampling reduces the dimensions by a factor of two in each direction after each pass. In this work, only eight levels of decomposition were used. Because the final two levels would consist of four-pixel and one-pixel images, respectively, which are basically useless for mammogram analysis, compared to the size of the entire breast. As a result, these levels are omitted from the wavelet analysis to speed calculation.
\begin{enumerate}[labelsep = .5em, leftmargin = 0pt, itemindent = 3em]
\item[i)] {Choice of Transform Methods:} Fig.~\ref{fig:comparison} shows the original mammogram and its four detail views obtained at the first decomposition level when the Db4 wavelet basis is used. It is shown that the wavelet maps have a lower resolution than the original image. Each view is sensitive to different features in the image. For example, the horizontal detail detects vertical changes in intensity, the vertical detail detects horizontal changes in intensity, the diagonal detail responds when the intensity is varying in both directions, and the approximation image is a low resolution version of the original image used as an input to the next coarser level of the decomposition. Fig.~\ref{fig:comparison TF} shows the Fourier transform view of the original mammogram. Compared with the wavelet maps, it can be seen that the wavelet transform provides multi-resolution decomposition, which means the wavelet maps at different levels reflect the image features of different sizes. Furthermore, spatial information is partially conserved. The wavelet maps in Fig.~\ref{fig:comparison} show the spatial distribution of information at particular size scales; in contrast, the Fourier transform would lose the spatial information and simply produce a map of the relative contributions of different frequencies over the entire image. This spatial information is useful for finding 
localized structures, such as microcalcifications and masses. These structures remain localized after the wavelet transform is applied, and they can then be distinguished from a more homogeneous background.

\begin{figure}[!htbp]
\centering
\subfigure{
\includegraphics[height = 2.92cm]{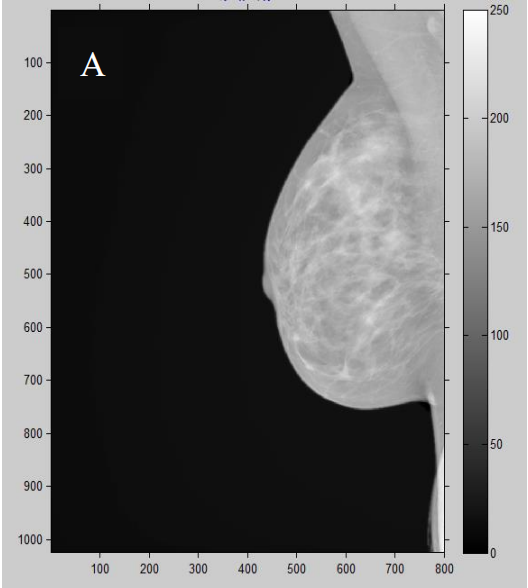}
}
\subfigure{
\includegraphics[height = 2.92cm]{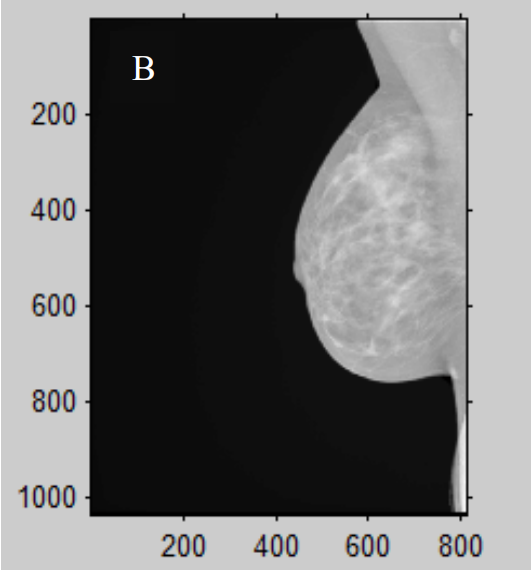}
}
\subfigure{
\includegraphics[height = 2.92cm]{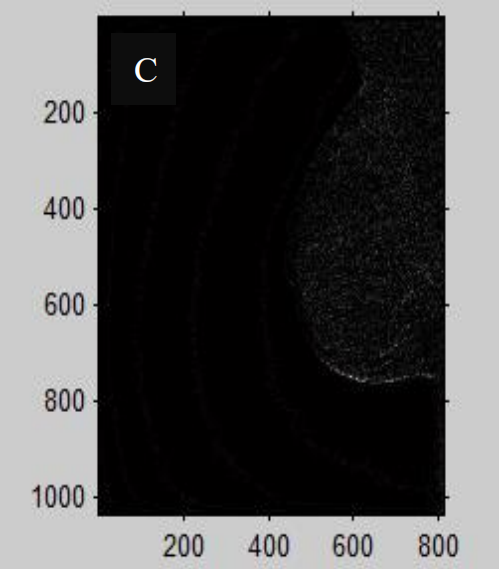}
}
\subfigure{
\includegraphics[height = 2.92cm]{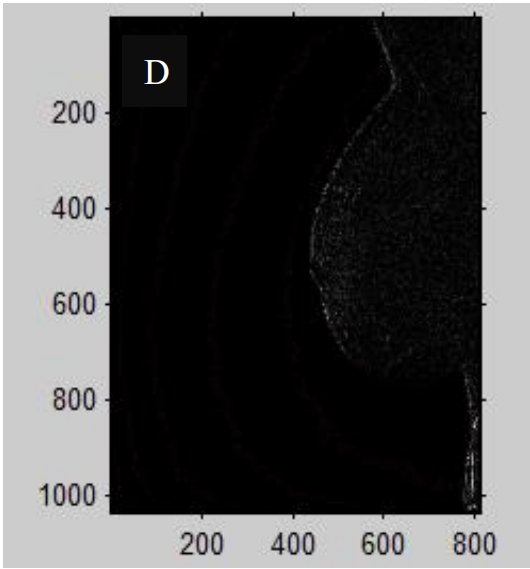}
}
\subfigure{
\includegraphics[height = 2.92cm]{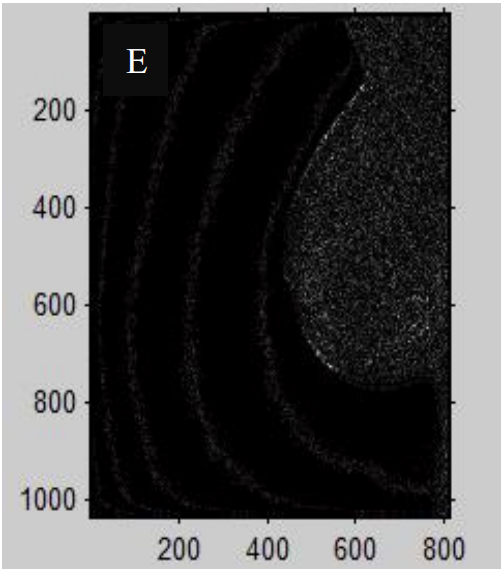}
}
\vspace{-0.6cm}
\caption[First level db4 wavelet decomposition: A. Original mammography image; B. Approximation view; C. Horizontal detail view; D. Vertical detail view; E. Diagonal view.]
{First level db4 wavelet decomposition: A. Original mammography image; B. Approximation view; C. Horizontal detail view; D. Vertical detail view; E. Diagonal view.}
\vspace{-0.4cm}
\label{fig:comparison}
\end{figure}

\begin{figure}[!htbp]
\centering
\includegraphics[height=2.95cm]{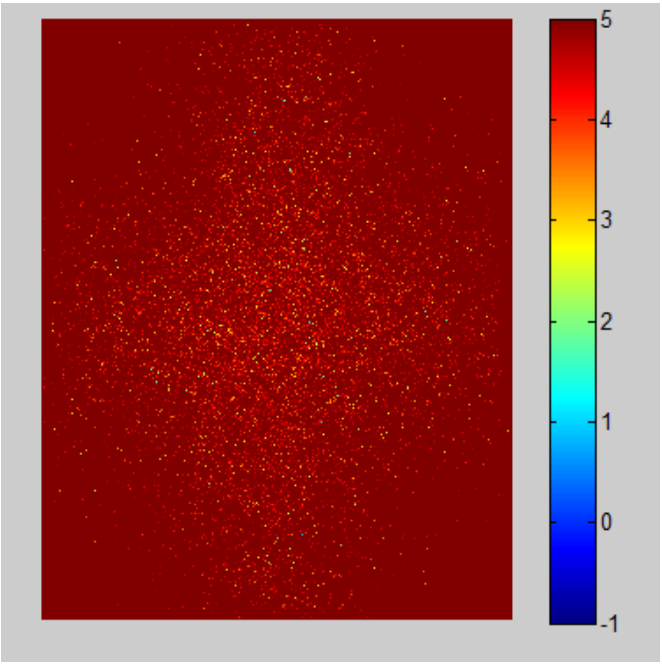}
\vspace{-0.3cm}
\caption[The Fourier transform view of the mammogram in Fig.~\ref{fig:comparison}.]
{The Fourier transform view of the mammogram in Fig.~\ref{fig:comparison}.}
\label{fig:comparison TF}
\vspace{-0.65cm}
\end{figure}

     \item[ii)] { Choice of Measurement:} in our experiment, four statistical features were extracted: mean intensity, standard deviation, skewness and kurtosis of the pixel intensities. Then, the mammogram analysis system uses some of these features to classify mammogram images as being normal or cancerous. 

\begin{itemize}[labelsep = .5em, leftmargin = 0pt, itemindent = 3em]
    \item {Mean:} the mean $\mu$ in this paper is obtained by calculating the average pixel value of the 
tissue region in the mammogram image. The equation is given by
\begin{equation}
    \mu = \frac{1}{N}\sum_{i,j}I(i,j)
\end{equation}
where $I(i,j)$ the pixel value at point $(i,j)$ of the mammogram image. N is the number of pixels in the tissue region of the image. The mean feature measures the average value of each detail views at different decomposition levels. Microcalcifications are usually tiny and bright. Compared with normal samples, microcalcifications have a slightly higher intensity in the high resolution maps. While masses are usually different in sizes and shapes, they could range from millimetres to several centimetres in width. Therefore, masses cannot be extracted from the background tissue through single scale or wavelet basis. However, masses are located in one region of tissue, and they are usually brighter than normal tissue. As a result, a 
slightly larger mean intensity can be measured through a wavelet basis, especially when different scales are used to detect masses.

\item{Standard Deviation:} the standard deviation $\sigma$, the estimate of the mean square deviation of grey pixel values, describes the dispersion of a local region. It is defined as
\begin{equation}
    \sigma = \sqrt{\frac{1}{N}\sum_{i,j}[I(i,j)-\mu]^2}
\end{equation}
It measures the variability in the brightness of the image over the tissue region. The value of the standard deviation would increase in the high spatial resolution levels of the wavelet map images that contain microcalcifications or masses, because they are brighter than normal parts of mammogram images.

\item{Skewness:} the third statistic feature measured from each wavelet map image is the skewness of the pixel intensities, which measures the degree of asymmetry. The skewness of a distribution of values is defined as the third central moment of the distribution, normalized by the cube of the standard deviation. It is given by
\begin{equation}
    S = \frac{1}{N}\sum_{i,j}[\frac{I(i,j)-\mu}{\sigma}]^3
\end{equation}
When a distribution has a larger right tail, then it shows a positive skewness. Even 
there is no significantly difference in the mean value or standard deviation, the 
skewness still changes because it is sensitive to the addition of a small number of unusually small or large values on a distribution.

\item{Kurtosis:} the fourth statistic measured from the wavelet maps is the kurtosis of the pixel 
intensities. The kurtosis of a distribution of values is defined as the fourth central 
moment of the distribution, normalized by the fourth power of the standard deviation 
of the distribution. The kurtosis K is given by
\begin{equation}
    K = \frac{1}{N}\sum_{i,j}[\frac{I(i,j)-\mu}{\sigma}]^4
\end{equation}
Kurtosis measures the narrowness of the central peak of a distribution compared with the size of the distribution’s tails. A distribution with a narrow peak and tails that drop off slowly has a large kurtosis compared with a distribution with a relatively wide peak but suppressed tails. The kurtosis and standard deviation of a distribution may be similar, but kurtosis is more sensitive to points distant from the mean than the standard deviation. Because of this, kurtosis is sensitive to the presence of microcalcifications and masses. It will rise when the number of unusual bright pixels increases in a wavelet map.

\end{itemize}
\end{enumerate}
\subsection{Feature Selection}

Since a large number of potential classification features are generated from each mammogram image, a selection process is needed to choose those features that are most effective at differentiating between normal and cancerous images. Specifically, there are four parameters measured from each wavelet map, with four wavelet maps per level and eight levels of decomposition. Thus, 16 features could be generated form each level of decomposition. To eliminate some of these, it was noted in N. Terki, etc.~\cite{b16} that peak signal to noise ratio (PSNR) improved when the level of decomposition
increases, and the image quality was better from third level of decomposition. Therefore, level 3 to level 8 of decomposition of the proposed three wavelet transform 
methods were applied in this work. In this case, 96 features would be generated from each of three wavelet transforms based on the 6 levels of wavelet decomposition.

Then, the generated 96 features from the wavelet transform were combined with the 6 features extracted from the Fourier transform. In other words, 3 different feature sets were created, and each of the feature sets contains features from one wavelet transform and the Fourier transform.

We adopted entropy-based feature selection in our work. The entropy of a variable X is defined as
\begin{equation}
\label{equ:entropy value}
    H(X) = -\sum_iP(x_i)\log_2(P(x_i))
\end{equation}
and the entropy of X after observing values of another variable Y is defined as
\begin{equation}
\label{equ: entropy}
    H(X | Y) = -\sum_jP(y_j)\sum_iP(x_i|y_j)\log_2(P(x_i|y_j))
\end{equation}
where $P(x_i)$ the prior probabilities for all values of X, and $P(x_i|y_j)$ is the posterior probabilities of X given the values of Y. The amount by which the entropy of X decreases reflects additional information about X provided by Y, and is called information gain~\cite{b11}, given by
\begin{equation}
\label{equ:gain}
    IG(X|Y) = H(X) - H(X|Y)
\end{equation}
If we have $IG(X|Y)>IG(Z|Y)$, it means a feature Y is regarded more correlated to feature X than to feature Z.

The entropy with the feature selection algorithm was implemented by the following steps:
\begin{itemize}[labelsep = .5em, leftmargin = 0pt, itemindent = 3em]
 \item[1)] Order features based on decreasing entropy values using \eqref{equ:entropy value}, and build a link list for all features;
\item[2)] Calculate the entropy of each feature in the link list related to the classification results using \eqref{equ: entropy};
\item[3)] Calculate the information gain of each feature using~\eqref{equ:gain} based on its two entropies obtained from step 1 and 2;
\item[4)] Compare each feature’s information gain with the next feature, and move the 
larger one ahead till the end of the link list;
\item[5)] Select the features with the information gain larger than the threshold set in the 
program.
\end{itemize}

In order to select the most effective features for differentiating between normal and cancerous mammogram images, less significant features are removed by entropy-based algorithm. This selection was achieved by sorting and selecting 
features with higher information gain values. The experimental results suggested that
the information gain of features from the db4-Fourier transform was higher than that 
of features from the bior6.8-Fourier transform, and the information gain of features 
from the db2-Fourier transform was the lowest among the three feature sets. In the 
features from the db2, db4, bior6.8, and Fourier transforms, we selected the top 12 
features (the optimal features) with their information gain values higher than 0.74.

\subsection{Image Classification}
In the final image classification stage, mammogram images were determined as either normal or cancerous based on the selected features. In proposed system, three classifiers (Linear Discriminant Analysis, Back-propagation Network, and Naive Bayes Classifier) were trained and tested. Moreover, combining the above mentioned classifiers (LDA, BP, NB), a voting classification scheme is further proposed for the mammogram analysis system in this research. In the voting classification scheme, where “1” represents cancerous mammograms from a classifier, and “0” represents normal mammograms.
When classifying a mammogram image, the voting classification decision is made by taking opinions of the majority of the three classifiers.
\begin{enumerate}[labelsep = .5em, leftmargin = 0pt, itemindent = 3em]
    \item[1)]  {Linear Discriminant Analysis:} the objective of LDA is to make the data points of different classes as far apart from each other as possible. In addition, it also aims at making the data points from the same class as close as possible. It can be implemented as:
\begin{itemize}[labelsep = .5em, leftmargin = 0pt, itemindent = 3em]
\item Constructing a matrix of feature vectors: all feature samples were read in as a matrix $X = \{x_1,x_2, ..., x_n\}$. Each feature data $x_i$ was regarded as a node i, and in the same way, another feature data $x_j$ was regarded as a node j. Node i and j were connected with a line if $x_i$ and $x_j$ were close, and they belonged to the same class.
\item Calculating scatter matrices: in this step, between-class scatter matrix $S_b$ and
within-class matrix $S_w$ were calculated using ~\eqref{equ:sb}~\eqref{equ:sw}.
\begin{align}
\label{equ:sb}
& S_b = \frac{1}{n}\sum_{i=1}^c(m^i-m)(m^i-m)^T \\
& S_w = \frac{1}{n}\sum_{i=1}^c(\sum_{j=1}^{n_i}(x_j^i-m^i)(x_j^i-m^i)^T) 
\label{equ:sw}
\end{align}

where $m$ is the total sample mean vector, $n_i$ is the number of samples in class $C_i$, $m^i$ is the average vector associated to $C_i$ class, $x_j^i$
is the $j-th$ sample vector in 
the $C_i-th$ class. $S_b$ and $S_w$ are named between-class scatter matrix and 
within-class matrix, respectively.

\item  LDA projection: data points $x_i$ were projected into the LDA subspace so that the matrix $S_w$ was non-singular. The transformation matrix of LDA was presented here as $W_{LDA}$. After projection, $S_b$ and $S_w$ became
\begin{align}
\centering
  &  \overline{S_b} = W_{LDA}^TS_bW_{LDA},\\
  &  \overline{S_w} = W_{LDA}^TS_wW_{LDA}
\end{align}
\item Computing the projection matrices: after adding the Lagrange multiplier and some derivation steps, the following function was achieved. It is also called the Fisher Linear Discrimination.
\begin{equation}
    S_w^{-1}S_bW = \lambda W
\end{equation}
It can be seen that $W$ is the eigenvector of matrix $S_w^{-1}S_b$.
\item Linear embedding: with the substitution of eigenvector, $W_{best}$ is easy to find 
by the following equation:
\begin{equation}
    W = S_w^{-1}(\mu_1 - \mu_2)
\end{equation}
where $\mu$ is the mean value (central point) of samples in each class.
\begin{equation}
    \mu_i = \frac{1}{n_i}\sum_{x\in C}x
\end{equation}
\end{itemize}
\item[2)] {Back-propagation Network:} the aim of the algorithm was to classify mammograms into two categories: 
cancerous or normal. Because the input features are 14 dimensional, and there are two kinds of mammograms to be classified, the construction of the BP network can be 
defined as “14-15, 2”. It means that there are 14 nodes in the input layer, 15 nodes in 
the hidden layer, and 2 nodes in the output layer. Furthermore, after random sorting of 
670 mammograms, 520 of them were randomly selected as the training dataset, the 
remaining 150 were chosen to test the classification performance of the BP network.
\begin{itemize}[labelsep = .5em, leftmargin = 0pt, itemindent = 3em]
    \item Network initialization: according to the input and desired output values (X and Y) of the network, we can set n nodes in the input layer, l nodes in the hidden layer, and m nodes in the output layer. The weight values ($w_{ij}$ and $w_{jk}$), the threshold value $a$ in the hidden layer, the threshold value $b$ in the output layer, the learning speed, and the activation functions should also be initialized.
\item Calculation of the hidden layer output: this output $H$ can be achieved through $X$, $w_{ij}$, and $a$.
\begin{equation}
    H_j = f(\sum_{i=1}^nw_{ij}x_i-a_j),\quad   j = 1,2,...l
\end{equation}
here, $l$ is the number of nodes in the hidden layer, $f$ is an activation function.
In this work, the activation function is chosen as
\begin{equation}
    f(x) = \frac{1}{1+e^x}
\end{equation}
\item  Calculation of the output layer output: O is determined through $H$,$w_{jk}$, and $b$.
\begin{equation}
    O_k = \sum_{j=1}^lH_jw_{jk}-b_k,\quad  k=1,2,...m
\end{equation}
 Calculation of error. Based on the output O and the estimated output Y, we can obtain the prediction error $e$. $e_k = Y_k - O_k$.
\item Weight update: the values can be updated using the following equations:
\begin{align}
\label{equ:w}
& w_{ij}=w_{ij}+\eta H_j(1-H_j)x(l)\sum_{k=1}^mw_{jk}e_k,j = 1,2,...l\\
& w_{jk} = w_{jk}+\eta H_je_k,\quad  j =1,2,...,l,k = 1,2,...m
\end{align}
in which, $\eta$ is the learning speed.
\item Threshold update: the threshold value can be updated using the following equations:
\begin{align}
& a_i = a_i+\eta H_j(1-H_j)\sum_{k=1}^mw_{jk}e_k,\quad  j=1,2,...l\\
& b_k = b_k+e_k,\quad k = 1,2,...m
\end{align}
    \item If the iteration is not over, the algorithm goes back to the second step.
\end{itemize}
\item[3)] {Naive Bayes Classifier:} suppose that there are m classes, $C$ = $\{C_1,C_2,...,C_m\}$. $A =\{A_1,A_2,...,A_n\}$ are the features for one dataset. Given an instance, its feature is $X$ = $\{X_1,X_2,...,X_n\}$, then the posterior probability that instance belongs to a class $C_i$ is $P$ = $P(C_i)$ = $(X|C_i)$. 
The Naive Bayes Classifier can be represented as $C(X)$ = $argmax(P(C_i)P(X|C_i)),{C_i\in C}$.
It indicates that the prediction accuracy reaches the maximum value when instance $X$ has the largest posterior probability. However, the posterior probability is difficult to calculate. 
Therefore, the ``Naive Bayes hypothesis'', which assumes all features $A_i$ are independent from each other, is introduced to the Naive Bayes Classifier. Thus, $P(A_i|C,A_j) = P(A_i|C)$, $\forall A_i,A_j,P(C)>0$.

The Naive Bayes classification algorithm can independently learn either the
conditional probability of each feature $A_i$
in the class $C(P(A_i|C))$, or the probability 
of each feature $A_i$. Replaced with a normalization factor ``$a$'', the posterior probability 
becomes
\begin{equation}
\begin{split}
   P(C = c|A_1=a_1...A_n=a_n)=&\\
  \alpha P(C=c)\prod\limits_{i=1}^nP(A_i|C=c)
\end{split}
\end{equation}
According to \eqref{equ:w}, the optimal classification ($C = C_i$) should satisfy
\begin{align}
\hspace{-2mm}& P(C_i|<a_1,...,a_n>)=\frac{P(<a_1,...,a_n>C_i)}{P(<a_1,...,a_n>)}P(C_i)\\
\hspace{-5mm}& P(C_i|<a_1,...,a_n>)>P(C_j|<a_1,...,a_n>),j\neq i
\end{align}
\end{enumerate}

\section{Experiments}
\label{experiments}
In this section, we conduct experiments to evaluate the performance of our system on mammogram image dataset. Since we adopted entropy-based feature selection method, we evaluated and compared the performance of three unique classifiers (LDA, BP, NB) using optimal and non-optimal features. Also, the classification performances of the three classifiers were compared with Receiver Operating Characteristic (ROC) curves. Furthermore, we compared the performance of voting scheme and unique classifier scheme (LDA, BP, NB).

\subsection{Dataset}
The dataset to be analyzed in this work were a gift from the Eastern Health in Newfoundland and Labrador of Canada. It consists of 1487 mammogram images, which is divided into a training set with 1040 mammogram images and a test set with 447 mammogram images. The images were all anonymous in the format of DICOM, which is a set of standard protocols in medical image processing, storage, printing, and transmission. They were authorized for our experiment by the Health Research Ethics Authority (HREA) in the reference number of 11312. All DICOM mammogram images were sampled to 1024×1024 pixels and reconfigured to PGM format.

\subsection{Implementation Details}
In this experiment, the programs of the image processing and the classification were developed in Matlab 2010b, and the program of feature selection was developed in Eclipse using JAVA. The computer used was based on Windows 10, Intel(R) Core(TM) CPU i7-8700 CPU @ 3.2GHz, 16GB RAM. 

\subsection{Evaluation Metrics}
The performance of a mammography screening system can be measured by two parameters: sensitivity and specificity. Sensitivity (true positive rate) is the proportion of the cases deemed abnormal when breast cancer is present. For example, if 100 women do have breast cancer among 1000 screened patients but only 90 are detected, then the sensitivity is 90/100 or 90\%. Sensitivity may depend on several factors, such
as lesion size, breast tissue density, and overallS image quality. In cancer screening protocols, sensitivity is deemed more important than specificity, because failure to diagnose breast cancer may result in serious health consequences for a patient. Almost fifty percent of cases in medical malpractice relate to “false-negative mammograms”~\cite{b21}.
Specificity (true negative fraction) is the proportion of cases deemed normal when breast cancer is absent. For example, if 100 cases of breast cancer are diagnosed in a set of 1000 patients, and the screening system finds 720 cases to be normal, the specificity is 720/900 or 80\%. Although the consequences of a false positive (diagnosing a normal patient as having breast cancer) are less severe than missing a 
positive diagnosis of cancer, specificity should also be as high as possible. False positive examinations can result in unnecessary follow-up examinations and procedures, and may lead to significant anxiety and concern for the patient.

\subsection{Results}
\begin{figure}[!htbp]
\centering
\subfigure{
\includegraphics[height = 2.92cm]{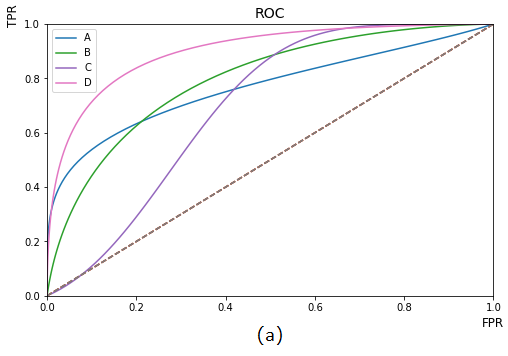}
}
\subfigure{
\includegraphics[height = 2.95cm]{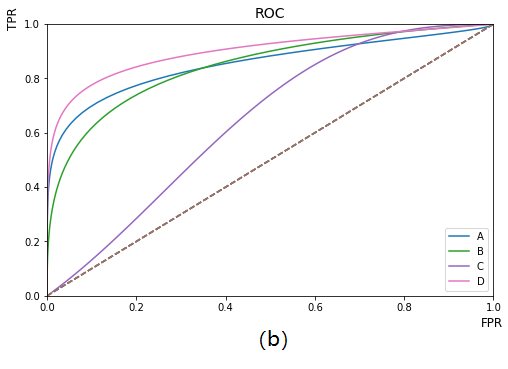}
}
\subfigure{
\includegraphics[height = 2.95cm]{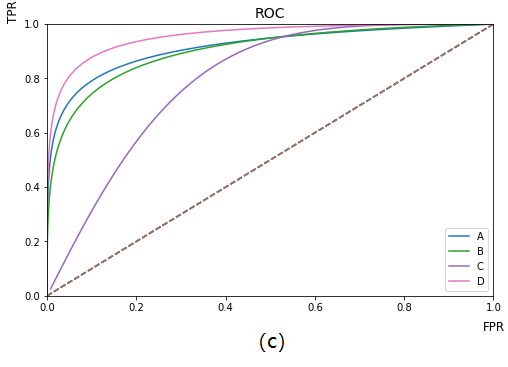}
}
\subfigure{
\includegraphics[height = 2.92cm]{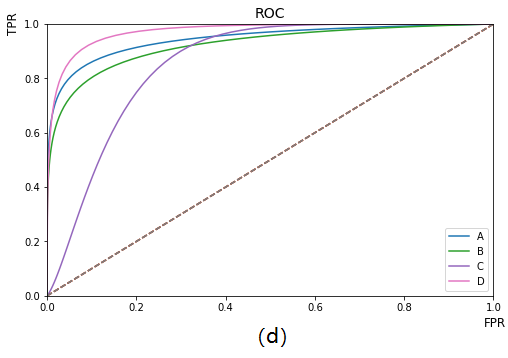}
}
\caption[ROC curves with the classifiers: A denoting LDA; B denoting BP; C denoting NB; D. Voting (a), 
(b), (c), and (d): performances of classifiers based on db2-Fourier, 
bior6.8-Fourier, db4-Fourier, and the optimal features respectively.]
{ROC curves with the classifiers: A. LDA; B. BP; C. NB; D. Voting (a), 
(b), (c), and (d): performances of classifiers based on db2-Fourier, 
bior6.8-Fourier, db4-Fourier, and the optimal features respectively.}
\label{fig: intensity matching}
\end{figure}

Table~\ref{classifier results} summarizes the accuracy, sensitivity and specificity performances of the three classifiers based on the selected features from db2-Fourier, bior6.8-Fourier, db4-Fourier, and the optimal features. Features including standard diversion, kurtosis and skewness extracted from the Fourier transform are obtained by the proposed entropy-based feature selection algorithm. By ranking the information gain, 12 optimal features with top information gain value for classification are selected. The Receiver Operating Characteristics curves were also plotted to facilitate comparison of the three classifiers as shown in Fig.~\ref{fig: intensity matching}.
\renewcommand{\arraystretch}{1.4}
\begin{table}[htbp]
\caption{Classification performances of three classifiers for the training dataset}
\begin{center}
\begin{tabular}{|p{1.0cm}|c|c|c|c|c|}
\hline
\multicolumn{2}{|c|}{\textbf{Classifier}}&{\textbf{LDA}}&{\textbf{BP}}&{\textbf{NB}}&{\textbf{VOTING}}\\ 
\hline
\multirow{3}{*}{\shortstack{db2-\\Fourier\\ Features}}&Accuracy & 80.69\% & 85.05\% & 81.03\% & 88.07\%\\
\cline{2-6}
 &Sensitivity&71.18\% & 83.05\% & 90.06\%&90.06\%\\
\cline{2-6}
 &Specificity & 89.03\% & 88.06\% & 70.08\%&92.03\%\\
\hline
\multirow{3}{*}{\shortstack{bior6.8-\\Fourier\\ Features}}&Accuracy & 81.01\% & 86.07\% & 83.03\%&89.01\%\\
\cline{2-6}
&Sensitivity&72.06\%& 84.65\% & 91.71\%&91.70\%\\
\cline{2-6}
&Specificity & 90.32\%& 89.01\%& 72.15\%&90.41\%\\
\hline
\multirow{3}{*}{\shortstack{db4-\\Fourier\\ Features}} & Accuracy & 84.03\% & 89.06\% & 86.02\% & 89.08\%\\
\cline{2-6}
&Sensitivity&74.24\%& 87.55\% &93.20\%&95.01\%\\
\cline{2-6}
&Specificity & 93.06\%& 92.05\%& 74.07\%&95.81\%\\
\hline
\multirow{3}{*}{\shortstack{The\\ optimal\\ Features}}&Accuracy & \textbf{88.02\%}& \textbf{94.14\%}& \textbf{89.83\%}&\textbf{96.06\%}\\
\cline{2-6}
&Sensitivity&\textbf{78.26\%}& \textbf{90.45\%}& \textbf{96.21\%}& \textbf{96.45\%}\\
\cline{2-6}
&Specificity & \textbf{96.88\%}& \textbf{95.03\%} & \textbf{80.60\%}&\textbf{97.45\%}\\
\hline
\end{tabular}
\label{classifier results}
\end{center}
\end{table}

According to the results in Table~\ref{classifier results}, classifiers achieved their highest classification performances using the optimal features, followed by the features selected from the db4 wavelet and Fourier transforms; whereas their performances are the lowest using features selected from the db2 wavelet and Fourier transforms. The specificity of the voting classifier with the optimal features is 1.64\% higher than that of the features of the db4-Fourier transform, 7.04\% higher than that of the bior6.8-Fourier transform, and 5.42\% higher than that of the db2-Fourier transform. Moreover, the specificity of the LDA, BP, NB classifiers with optimal features is better than that of the three feature methods, i.e., the db4-Fourier, db2-Fourier, and bior6.8-Fourier transform.

In addition, it is also clear that the voting classification scheme outperforms each individual base classifier. The reason that optimal features achieve the highest performance could be the information gain of the optimal features is the highest among the four different feature sets. In other words, the features in the optimal feature set are more correlated to the mammogram class than any of the other features.

What's more, it can be found from Table~\ref{classifier results} and ROC curves that the voting scheme achieves the highest accuracy using the default parameters in the proposed mammogram analysis system.  
On the other hand, the Naive Bayes Classifier among the three single classifiers (LDA, BP, NB) achieves the highest specificity, and the Back Propagation network achieves the highest accuracy based on all the four feature sets. This result suggests that among the three classifiers, the NB classifier is more sensitive to classify cancerous mammograms, the LDA classifier gives better classification in normal mammograms, and the BP neural network works well in both of normal and cancerous mammograms. The voting classification scheme performs best.

\section{Conclusions}
\label{conclusions}
 In this study, a computer-aided automatic mammogram analysis system, which consists of image processing, feature selection and image classification, is proposed to improve the detection performances. In conclusion, the experimental results on mammogram image dataset from the Eastern  Health in Newfoundland and  Labrador of Canada demonstrated that the proposed automatic mammogram analysis system could effectively improve the classification performances, especially using the voting classification scheme based on the selected optimal features. In the future, more wavelet bases can be explored to extract valuable features with high information gain.

\vspace{12pt}

\begin{thebibliography}{00}
\bibitem{b1} National Cancer Institute of Canada (2004). Canadian Cancer Statistics 2004.
Toronto, Canada.
\bibitem{b2} American Cancer Society (2014). Global Cancer Facts $\&$ Figures. Retrieved from http://www.cancer.org/acs/groups/content/@research/documents/webcon\\
tent/acspc-042151.pdf.
\bibitem{b3}  Canadian Cancer Society (2013). Breast Cancer Statistics at a Glance. Retrieved
from:http://www.cancer.ca/en/cancer-information/cancer-type/breast/statistics/?region=nl
\bibitem{b4} R. L. Birdwell (2009). The preponderance of evidence supports computer-aided detection for screening mammography. Radiology,253(1), 9-15.
\bibitem{b5} S. Wu, et al. (1996). Application research of laser scanning microscopy for early diagnosis of tumors. Proceedings of SPIE, 2887, 190-192.
\bibitem{b6} X. Kang (2001). Modern Medical Imaging Technology. Tianjing: Tianjing Technology Translation Publication Company.
\bibitem{b7} A. Kusiak (2001). Feature transformation methods in data mining. IEEE Trans.
Electron. Packag. Manuf. 24(3), 214-221.
\bibitem{b8} Discrete Fourier Transform (DFT) (2004). Retrieved from http://home.eng.iastate.edu/~julied/classes/ee524/LectureNotes/l5.pdf .
\bibitem{b9} Q. Feng, Z. Yang (2000). Practical Wavelet Analysis. Xi’an: Xi’an electronic 
science and technology university press.
\bibitem{b10}  I. Inza, P. Larranaga, R. Blanco, A.J. Cerrolaza (2004). Filter versus wrapper 
gene selection approaches in DNA microarray domains. Artif. Intell. Med. 31, 91–103.
\bibitem{b11} M.G.E. Schneiders (2001). Wavelets in Control Engineering. Master’s thesis,Eindhoven University of Technology.
\bibitem{b12} N. Otsu (1979). A threshold selection method from gray-level histograms. IEEE 
Trans. Sys. Man. Cyber. 9 (1), 62–66.
\bibitem{b13} H. Liu, L. Yu (2005). Toward integrating feature selection algorithms for 
classification and clustering. IEEE Trans. Knowl. Data Eng., 17(4), 491-502.
\bibitem{b14} Y. Saeys, et al. (2007). Bioinformatics (2007). A review of feature selection techniques in bioinformatics. Bioinformatics, 23 (19), 2507-2517.
\bibitem{b15} Y. Lei, L. Huan (2003). Feature selection for high-dimensional data: a fast 
correlation-based filter solution. Proc. Twentieth International Conference on Machine Learning (ICML-2003).
\bibitem{b16} N. Terki, N. Doghmane, A. Ouafi, Z. Baarir (2004). Study of effect of filters and
decomposition level in wavelet image compression.
\bibitem{b17} Hala Al-Shamlan and Ali El-Zaart, “Feature Extraction Values for Breast Cancer Mammography Images”, International Conference on Bioinformatics and Biomedical Technology 2010.
\bibitem{b18} D. C. Hope E. Munday S. L. Smith., “Evolutionary Algorithms in the Classification of Mammograms”, Proceedings of the 2007 IEEE Symposium on Computational Intelligence in Image and Signal Processing (CIISP 2007).
\bibitem {b19} Shahin, Osama R and Attiya, Gamal.  ``Classification of mammograms tumors using fourier analysis'', (IJCSNS 2014). 
\bibitem{b20} E.J. Kendall, M. Barnett, and K. Chytyk-Praznik (2013). Automatic detection of anomalies in screening mammograms. BMC Medical Imaging, 13-43.
\bibitem{b21} Physician Insurers Association of America: Breast Cancer Study (1995). 
Washington, DC.
\end{thebibliography}
\end{document}